\documentclass[a4paper]{jpconf}
\usepackage{graphicx}
\begin{document}
\title{The bursting of housing bubble as jamming phase transition}

\author{Katsuhiro Nishinari$^{1}$, Mitsuru Iwamura$^{2}$, Yukiko Umeno
Saito$^{3}$ and Tsutomu Watanabe$^{4}$}

\address{
$^{1}$ Research Center for Advanced Science \& Technology,University of Tokyo,
and PRESTO, Japan Science and Technology Corporation}
\address{
$^{2}$ Waseda Business School, Waseda Univeristy}
\address{
$^{3}$ Economic Research Center, Fujitsu Research Institute, and
Hitotsubashi University}
\address{
$^{4}$ Hitotsubashi University and Canon Institute for Global Studies}

\ead{tknishi@mail.ecc.u-tokyo.ac.jp}

\begin{abstract}
In this paper, we have proposed a bubble burst model by focusing on
 transaction volume incorporating
 a traffic model that represents spontaneous traffic jam.
 We find that the phenomenon of bubble burst shares many similar
properties with traffic jam formation on highway by comparing data
taken from the U.S. housing market.
Our result suggests that transaction volume could be
a driving force of bursting phenomenon.
\end{abstract}

%%%%%%%%%%%%%%%%%%%%%%%%%%%%%%%%%%%%%%%%%%%%%%%%%%%%%%%%%%%%%%%%%%%%
\section{Introduction}
\label{sec1}
%%%%%%%%%%%%%%%%%%%%%%%%%%%%%%%%%%%%%%%%%%%%%%%%%%%%%%%%%%%%%%%%%%%%
Fluctuations in real estate prices have substantial impacts on
economic activities. For example, land prices in Japan exhibited
a sharp rise in the latter half of the 1980s, and its rapid reversal
in the early 1990s. This large swing had led to a significant
deterioration of the balance sheets of firms, especially those of
financial firms, thereby causing a decade-long stagnation of the
Japanese economy, which is called Japan's ``lost decade''.
A more recent example is the U.S. housing market bubble, which
started somewhere around 2000 and is now in the middle of collapsing.
This has already caused substantial damages to financial systems
in the U.S. and the Euro area, and it is expected that it may
spread worldwide as in the case of the Great Depression in the 1920s and 30s.

These recent episodes have rekindled researchers' interest on the
issue of bubbles. Economists have been regarding this phenomenon
as a disorder in prices. Specifically, they define bubbles as a
temporary deviation of asset prices from the corresponding fundamental
values, which are basically determined by investors' expectations
about future dividend stream and appropriate discount rates.
Interestingly, even researchers in other areas, such as econophysics,
share this empirical strategy in the sense that they look for
abnormal behaviors in prices\cite{ZS,WT}.
However, this research strategy has overlooked an important aspect
of bubbles; namely, fluctuations in asset prices tend to be
closely correlated with those in the volume of transactions.

Fig.1 depicts fluctuations in housing prices and the volume of
house transactions in the U.S., which shows a positive correlation
between the two variables over the business cycles.
%%%%%%%%%%%%%%%%%%%%%%%%%%%%%%%%%%%%%%%
\begin{figure}
 \begin{center}
\resizebox{0.65\textwidth}{!}{%
  \includegraphics{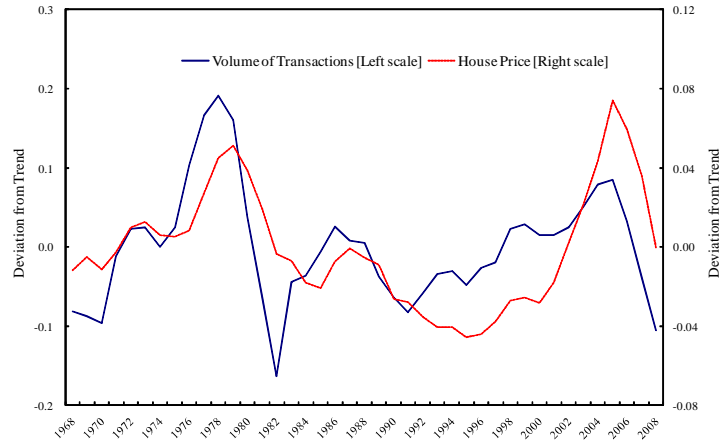}
}
\caption{Fluctuations in price and
 transaction volume in the U.S.
 single-family house market. Both variables represent deviations from a linear trend.
 }
\label{fig1}
  \end{center}
\end{figure}
%%%%%%%%%%%%%%%%%%%%%%%%%%%%%%%%%%%%%%%
More importantly, it shows that a change in transaction volume
tends to lead a change
in prices by one year or two\cite{St,WL}. It is reported that similar
relationships were observed for other real estate markets in other
countries, including Japan, U.K., Hong Kong, and
Singapore\cite{MR,AM,LL,EL}.
These evidences suggest that some sort of
interaction between prices and the volume of transactions plays an
important role in the process of bubble and its bursting\cite{St,GM,CL},
and that
fluctuations in transaction volume, rather than those in prices,
could be its driving force. Given this understanding, we focus more
on transaction volume in this paper, and seek to propose a model
which explains an emergence of temporary deviation of transaction
volume from its appropriate level, as well as its reversal.

%%%%%%%%%%%%%%%%%%%%%%%%%%%%%%%%%%%%%%%%%%%%%%%%%%%%%%%%%%%%%%%%%%%%
\section{Traffic jam perspective}
\label{sec2}
%%%%%%%%%%%%%%%%%%%%%%%%%%%%%%%%%%%%%%%%%%%%%%%%%%%%%%%%%%%%%%%%%%%%
The idea of our modeling stems from traffic jam on highway.
Each vehicle on highway usually tries to move faster if
there is enough space in front of it.
However, such flow becomes unstable and soon changes into congested state
if the density of vehicles becomes large enough.
Fig.2 is a typical
observed data of flow-density relation of vehicles on highway, which
is called the {\it fundamental diagram}. Note that flow is defined
as the multiplication of density and average velocity, and density is the
percentage of the occupied area of vehicles divided by the road area.
%%%%%%%%%%%%%%%%%%%%%%%%%%%%%%%%%%%%%%%
\begin{figure}[h]
 \begin{center}
\resizebox{0.6\textwidth}{!}{%
  \includegraphics{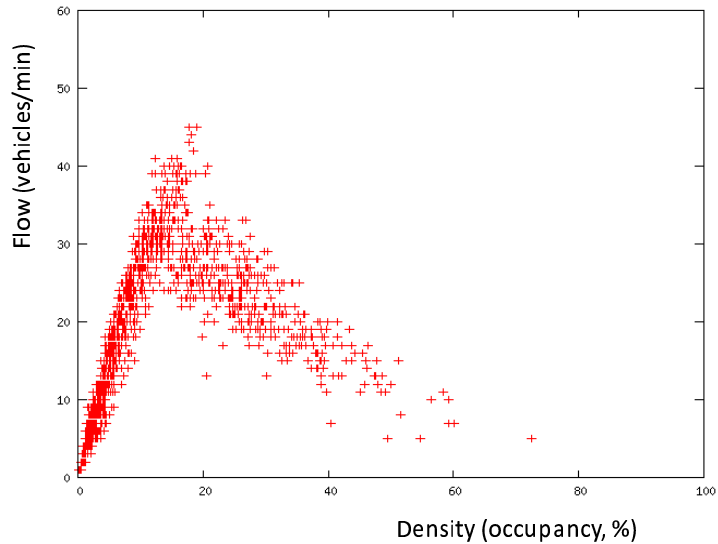}
}
\caption{Typical flow-density diagram
 taken from Tokyo Metropolitan highway. Below (above) the critical density
 flow increases (decreases) according to the increase of density. Around the
 critical density $15\%-20\%$ we see metastable state which corresponds
 unstable free flow.}
\label{fig2}
 \end{center} 
\end{figure}
%%%%%%%%%%%%%%%%%%%%%%%%%%%%%%%%%%%%%%%

From this figure we see that flow increases according to
the increase of density up to 15\%, which corresponds to free flow.
Instead, flow decreases above the density 20\%, which correspond to jam flow.
Thus we have phase transition
from free to jam state at critical density 15\%, and we find that
free flow overlaps jam flow between the range 15\%-20\% in the diagram.
A schematic picture of this diagram is given in Fig.3.
%%%%%%%%%%%%%%%%%%%%%%%%%%%%%%%%%%%%%%%
\begin{figure}[h]
 \begin{center}
\resizebox{0.6\textwidth}{!}{%
  \includegraphics{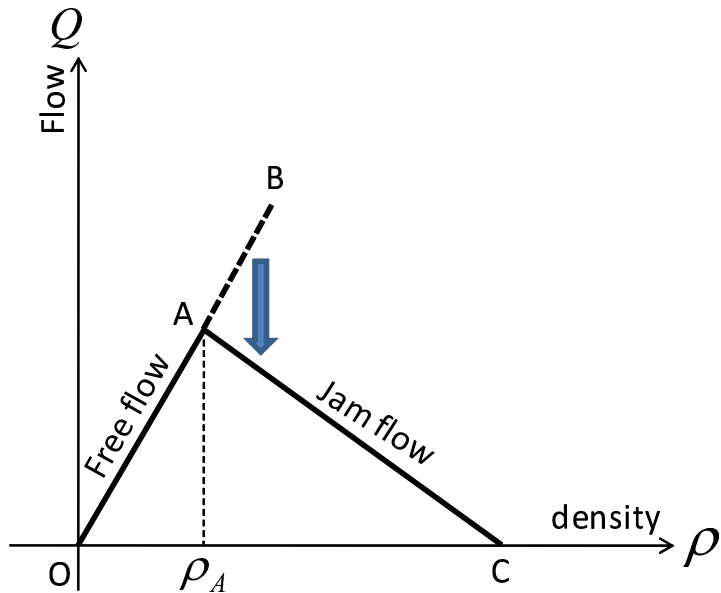}
}
\caption{Schematic flow-density diagram of observed data. There is
 metastable state $A-B$ above the critical density $\rho_A$.}
\label{fig3}
  \end{center}
\end{figure}
%%%%%%%%%%%%%%%%%%%%%%%%%%%%%%%%%%%%%%%

Free flow on $A-B$ is dangerous and unstable,
since the average velocity of vehicles is high
although the headway between successive vehicles is small.
Thus small perturbation, e.g., weak brakes or slow down due to slope
is enough to change the free flow into jamming state. Thus the branch
$A-B$ is called metastable state, whose lifespan is less than 10
minutes in reality and it suddenly changes into jam flow.
This scenario is indicated by an arrow in Fig.\ref{fig3}.
The existence of the metastable state is crucial for accounting
so-called phantom jam, i.e., traffic jam on highway without bottlenecks.
Drivers' tendency for moving faster will lead to small headway
with high velocity, which eventually causes metastable flow.
Then due to small perturbation traffic jam is
endogenously formed  spontaneously.
This has been confirmed by a simple model
called the slow-to-start (SlS) model\cite{TT,NT},
as well as an experiment\cite{NJP}.

Now we come to the point that
there is similarity between spontaneous jam formation and bubble burst.
Traffic jam occurs because there are too many cars
relative to the available road space, while
``transaction jam'' in a housing market occurs
because there are too many houses traded in the market relative to the total amount of liquidity (or money) supplied to the market by investors, including banks, who are outside the market.
Transaction is an exchange of a house and money, and vehicle's motion
is also considered as an exchange of a vehicle and a free space in front
of it. Thus we propose the following analogy: For vehicle traffic,
\begin{eqnarray}
{\rm Density}&=&
 \frac{\rm Space\,\, for\,\, cars}
 {\rm Total\,\, available\,\, space},\nonumber\\
{\rm Average\,\, velocity}&=&
 \frac{\rm Number\,\, of\,\, moved\,\, cars}{\rm Number\,\, of\,\, cars},
 \nonumber
\end{eqnarray}
and for housing market transactions,
\begin{eqnarray}
{\rm Density}&=&
 \frac{\rm Housing\,\, inventory}{\rm Total\,\, available\,\, liquidity},\\
{\rm Average\,\, velocity}&=&
\frac{\rm Transactions}{\rm Housing\,\, inventory},
\end{eqnarray}
where housing inventory and transactions are evaluated as USD or other monetary unit.

Furthermore, we conjecture that
metastable state plays a great role also in bubbule burst, as so in
phantom jam.
More precisely, we think that before the burst occurs,
there is metastable transaction between real estate companies.
We will show such modeling below, and compare the result with
observed data taken from the US housing market.

%%%%%%%%%%%%%%%%%%%%%%%%%%%%%%%%%%%%%%%%%%%%%%%%%%%%%%%%%%%%%%%%%%%%
\section{Bubble burst model}
\label{sec3}
%%%%%%%%%%%%%%%%%%%%%%%%%%%%%%%%%%%%%%%%%%%%%%%%%%%%%%%%%%%%%%%%%%%%
%%%%%%%%%%%%%%%%%%%%%%%%%%%%%%%%%%%
\subsection{Transaction and price}
\label{sec31}
\begin{figure}[h]
\begin{center}
 \resizebox{0.65\textwidth}{!}{%
  \includegraphics{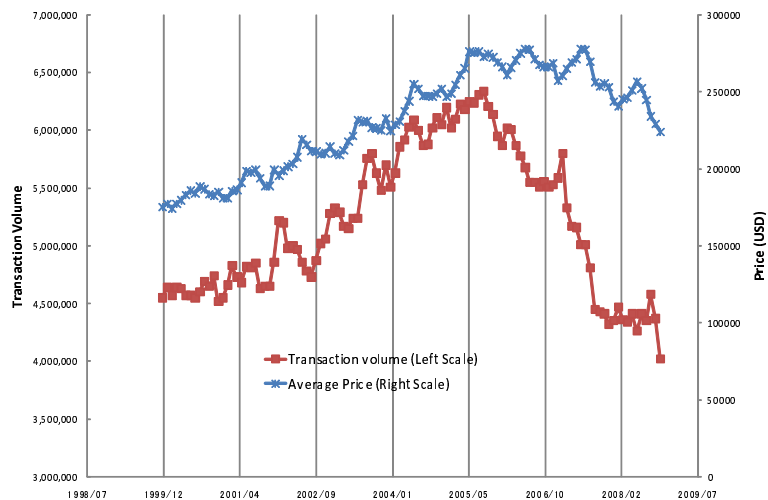}
}
\caption{Monthly fluctuations in average house prices and
 transaction volumes in the U.S. housing market
 from Dec 1999 to Nov 2008.}
\label{fig4}
 \end{center}
\end{figure}

%%%%%%%%%%%%%%%%%%%%%%%%%%%%%%%%%%%
Before considering a model for bubble burst, let us focus on the
housing sales from December 1999 to November 2008 in the U.S., during which
both transaction volumes and prices showed a significant swing in our dataset.
From fig.\ref{fig4}, we immediately see
 that a decline in transaction volume occurred earlier than a fall in price. We also find
 that housing price is almost saturated around the bursting period,
 which is also clearly seen in the data of Japan's bubble burst in 1992.
 Thus in our model we focus on transaction volume instead of price in order
 to grasp the essence of the bursting phenomena.
%%%%%%%%%%%%%%%%%%%%%%%%%%%%%%%%%%%%%%%
%%%%%%%%%%%%%%%%%%%%%%%%%%%%%%%%%%%%%%%
%%%%%%%%%%%%%%%%%%%%%%%%%%%%%%%%%%%
\subsection{Model description}
\label{sec32}
\begin{figure}[h]
 \begin{center}
\resizebox{0.6\textwidth}{!}{%
  \includegraphics{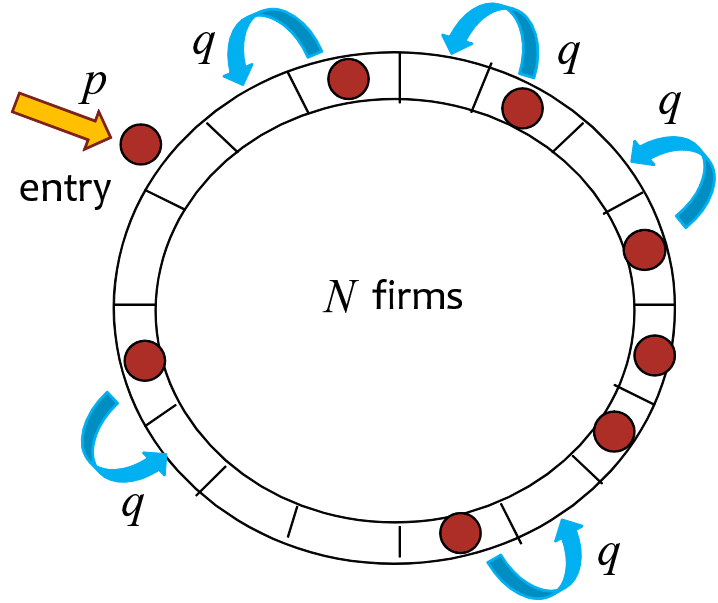}
}
\caption{Bubble burst model on a circular market. There are
 $N$ firms dealing houses represented by circles.
 The probability of successful transaction is
 given by $q$.}
\label{fig5}
  \end{center}
\end{figure}

%%%%%%%%%%%%%%%%%%%%%%%%%%%%%%%%%%%
Now we propose a simple model for bubble burst incorporating the
SlS model.
Consider an economy with symmetric $N$ firms
dealing with houses, which are identified by $i$ ($i=1,\cdots,N$)
and located along a circle (Fig.\ref{fig5}).
%%%%%%%%%%%%%%%%%%%%%%%%%%%%%%%%%%%%%%%
%%%%%%%%%%%%%%%%%%%%%%%%%%%%%%%%%%%%%%%
Firm $i$ is allowed to purchases a house from firm $i-1$,
but not allowed to do so from any other firm. For simplicity,
it is assumed that the amount of cash each firm owns before
buying a house is identical to the price of a house,
so that each firm has its entire asset either in the form of
cash or in the form of a house.

\begin{figure}[h]
 \begin{center}
\resizebox{0.5\textwidth}{!}{%
  \includegraphics{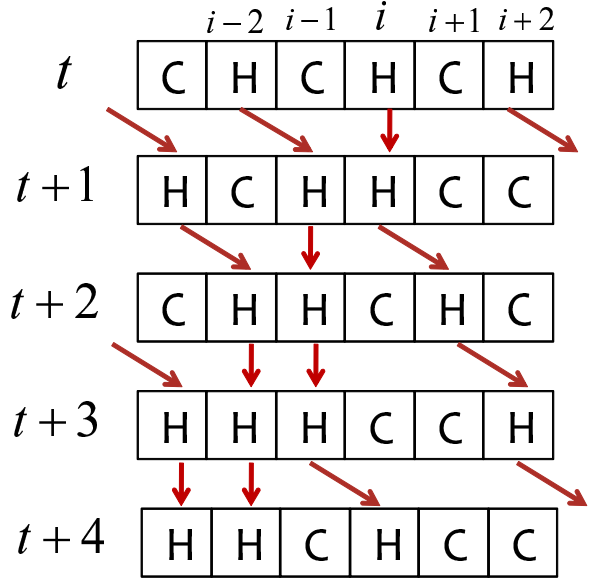}
}
\caption{Hesitation of firm $i$ in period $t+2$, which is due to
 the collapse of credit in the period $t$, 
 leads to jam of transaction in this economy. H and C
 represent house and cash, respectively.}
\label{fig55}
  \end{center}
\end{figure}

There are two rules governing transactions in this economy.
First, all transactions must be in the form of an exchange of
cash and a house, and no barter transactions (i.e. transactions
between a house and another house) are allowed. This implies that transaction
between firms $i$ and $i-1$ never takes place in period $t$ unless firm
$i$ holds cash at the beginning of that period.
This rule corresponds to what macroeconomists call ``cash-in-advance''
constraint. Second, firm $i$ becomes timid if firm $i$ fails to make a
transaction with firm $i+1$ in period $t$. Specifically, firm $i$
hesitates to deal with firm $i-1$ and
refuses to purchase a house in period $t+2$ even if
firm $i$ successfully has sold a house to firm $i+1$ in period $t+1$,
therefore holding cash at the beginning of period $t+2$ (Fig.\ref{fig55}).
This is because, ceteris paribus, firm $i$ is able to increase
the probability that firms behind him (firm $i+1, i+2, i+3,\cdots$) hold
cash, instead of a house, thereby reducing the probability that
he will be refused to sell a house to firm $i+1$.
This hesitation rule represents firms' preference to cash as a means to
store value until the next period, because of its general acceptability.
%%%%%%%%%%%%%%%%%%%%%%%%%%%%%%%%%%%%%%%
%%%%%%%%%%%%%%%%%%%%%%%%%%%%%%%%%%%%%%%

In the modeling of vehicle traffic, there are also two important rules:
the exclusion principle and the SlS rule which correspond to
the above two rules in our model. Firms and houses correspond to
 the road cells and vehicles, respectively. The exclusion principle in
 vehicle traffic is that  a cell can accommodate at most one vehicle for
 avoid collision. The SlS rule represents an inertia effect of vehicles,
 i.e., once a vehicle stops, then it waits extra one time step to move forward
 after the front cell becomes empty. In our model, this wait corresponds
 to the hesitation to make another transaction due to the lack of confidence about the trade partner's ability to pay, and thus the fear of default. Moreover, this hesitation rule is crucial to have
 chain reaction of transaction fails, which leads to the bubble burst. This
 is similar to chain reaction
 of brakes on a highway under the SlS rule, which is generally
 observed in real traffic data.

%%%%%%%%%%%%%%%%%%%%%%%%%%%%%%%%%%%
\subsection{Simulations}
\label{sec33}
%%%%%%%%%%%%%%%%%%%%%%%%%%%%%%%%%%%
Given the above rules, we conduct simulations.
We assume that there exists no housing inventory in this market at
the beginning of period 0, but in each period, firm $i$,
which is randomly chosen, purchases a new house from someone
outside this market with probability $p$ if (1) firm $i$ does not hold
a house (and therefore owns cash), and (2) firms ahead and
behind firm $i$ (namely, firms $i-1$ and $i+1$) do not hold a house either.
In words, firm $i$ wants to purchase a house from firm $i-1$,
but cannot do that because firm $i-1$ does not own a house.
At the same time, firm $i$ expects that firm $i+1$ will be able to
buy a house from him because firm $i$ owns cash.
It is only in this situation that firm $i$ brings in a new house to
the market from outside. We also assume that the probability of
successful transaction between any two adjacent firms
(the one with cash and the one ahead with a house) is given by $q$ (Fig.\ref {fig5}).

\begin{figure}[h]
 \begin{center}
\resizebox{0.45\textwidth}{!}{%
  \includegraphics{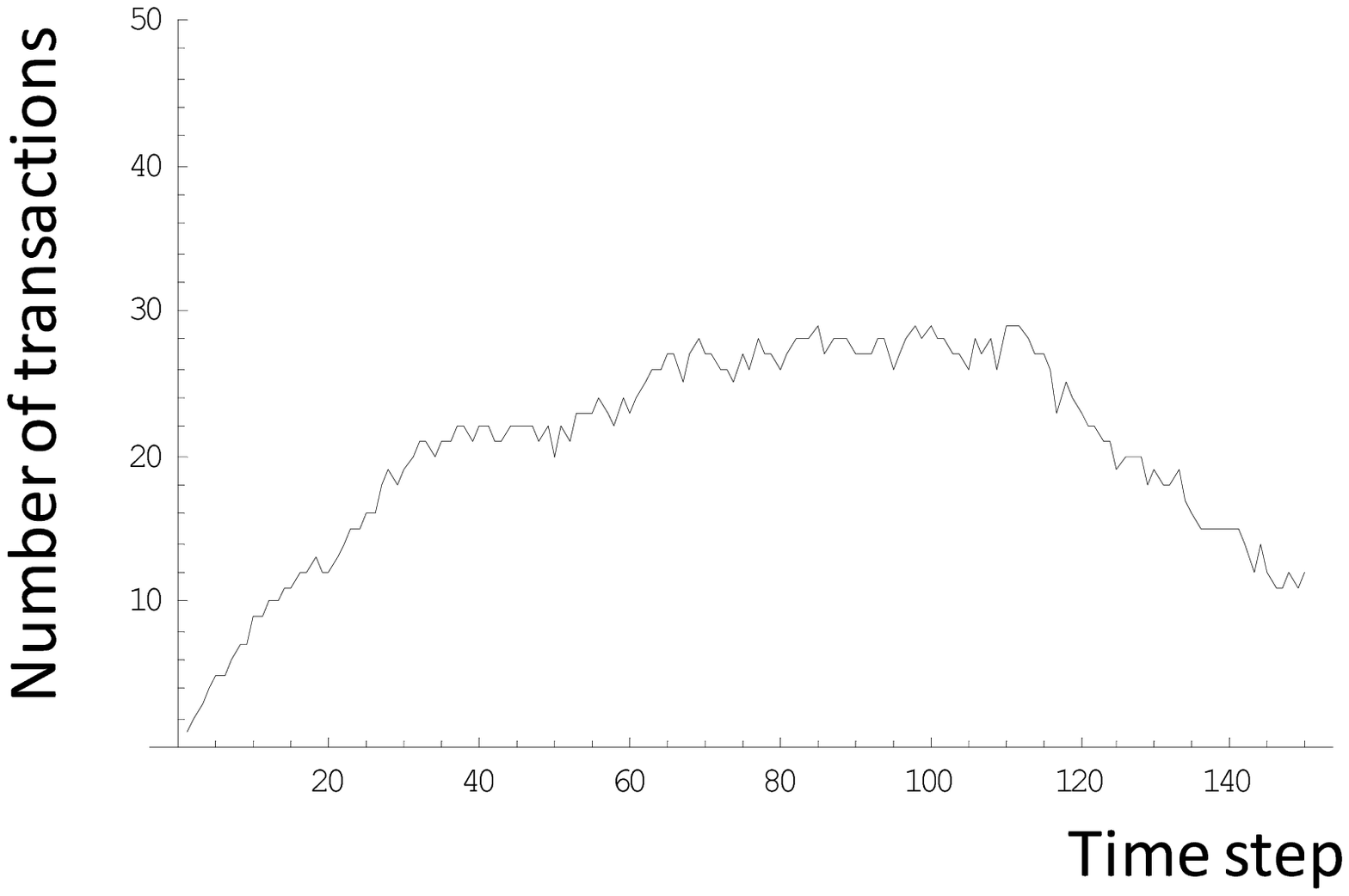}
}
\resizebox{0.45\textwidth}{!}{%
  \includegraphics{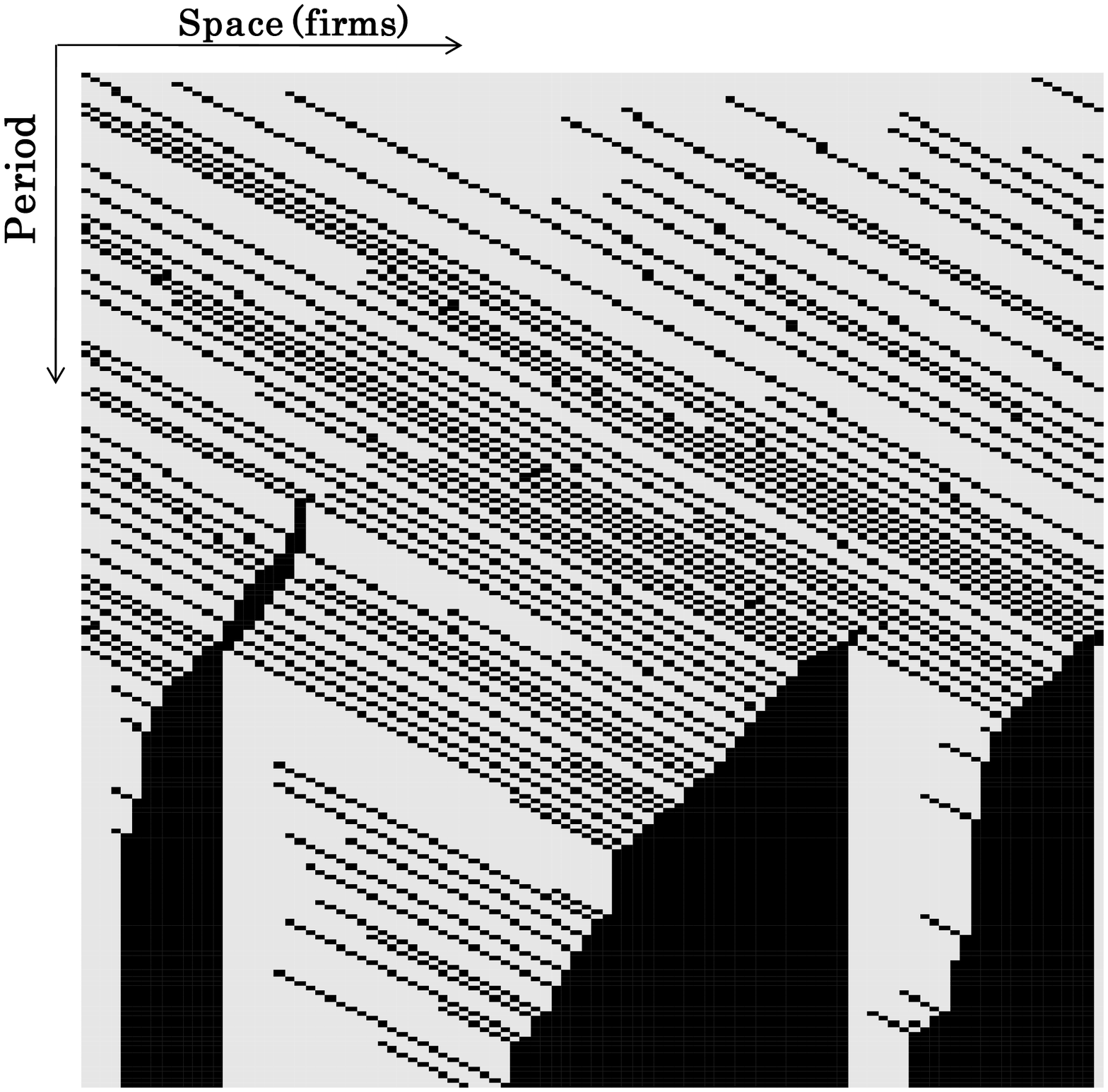}
}
\caption{(left) Fluctuations in transaction volume over time.
(right) Corresponding spatiotemporal figure of transaction. Small black
 squares represent houses which move to right due to the success of
 deal. Parameters N and q are set at N=100 and q=0.99 in the simulation.}
\label{fig6}
  \end{center}
\end{figure}

Fig.\ref{fig6} (left) shows fluctuations in transaction volume over time.
An important thing to note is that transaction volume exhibits
an abrupt decline in $t$=115 after a quiet
period in which transaction volume is kept fairly stable.
This abrupt decline is an endogenous event, and can be seen
as phase transition phenomenon emerging from metastable state.
Fig.\ref{fig6} (right) shows an example of spatiotemporal figure of
transaction flow. We see jam of transaction
emerges and rapidly grows endogenously,
which is very similar to an abrupt
change from free to congested flow in highway traffic.
%%%%%%%%%%%%%%%%%%%%%%%%%%%%%%%%%%%%%%%
%%%%%%%%%%%%%%%%%%%%%%%%%%%%%%%%%%%%%%%

\begin{figure}[h]
 \begin{center}
\resizebox{0.65\textwidth}{!}{%
  \includegraphics{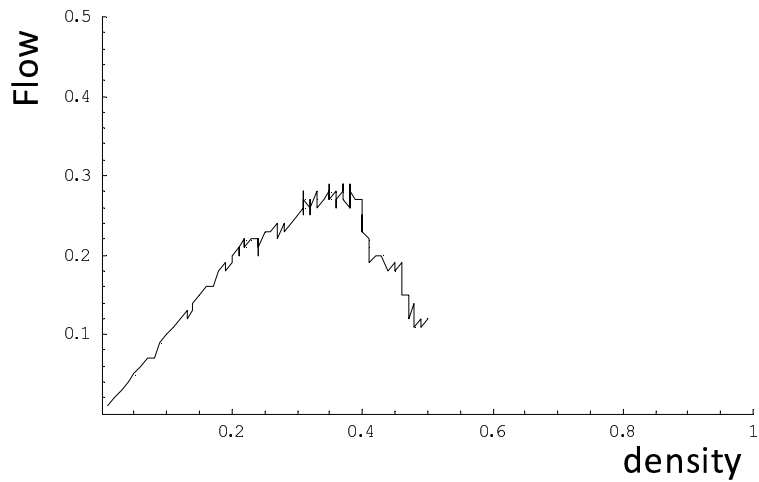}
}
\caption{Flow-density diagram 
Parameters N and q are set at N=100 and q=0.99.}
\label{fig7}
  \end{center}
\end{figure}

The fundamental diagram is given in Fig.\ref{fig7}. Flow is defined by
multiplying eq.(1) and eq.(2).
Due to the entry of
houses we see linear increase of flow as density increases.
The critical density is about 1/3 in this model \cite{SNI}, so we see
the metastable transaction between 0.3 and 0.4.
Then this market suddenly breaks due to the collapse of credit,
and bubble bursts.

%%%%%%%%%%%%%%%%%%%%%%%%%%%%%%%%%%%%%%%
%%%%%%%%%%%%%%%%%%%%%%%%%%%%%%%%%%%%%%%

%%%%%%%%%%%%%%%%%%%%%%%%%%%%%%%%%%%%%%%%%%%%%%%%%%%%%%%%%%%%%%%%%%%%
\subsection{Comparison with data}
\label{sec34}
\begin{figure}[h]
 \begin{center}
\resizebox{0.65\textwidth}{!}{%
  \includegraphics{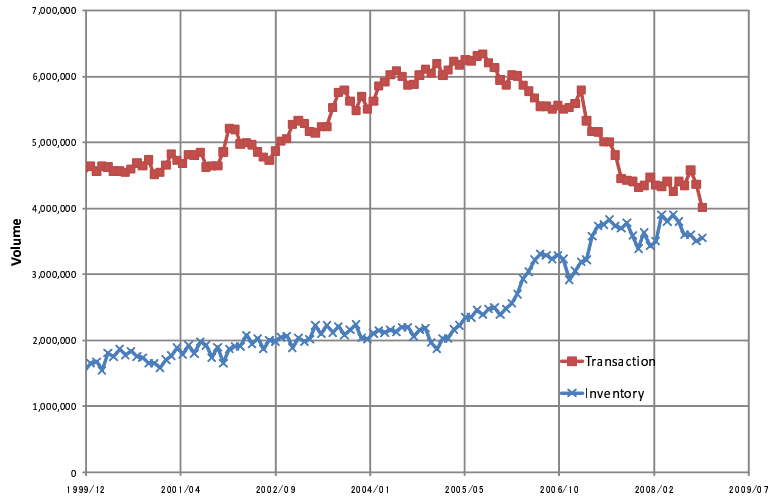}
}
\caption{Transaction volume and
 inventory in the U.S. housing market
 from Dec 1999 to Nov 2008.}
\label{fig8}
  \end{center}
\end{figure}

\begin{figure}[h]
 \begin{center}
\resizebox{0.49\textwidth}{!}{%
  \includegraphics{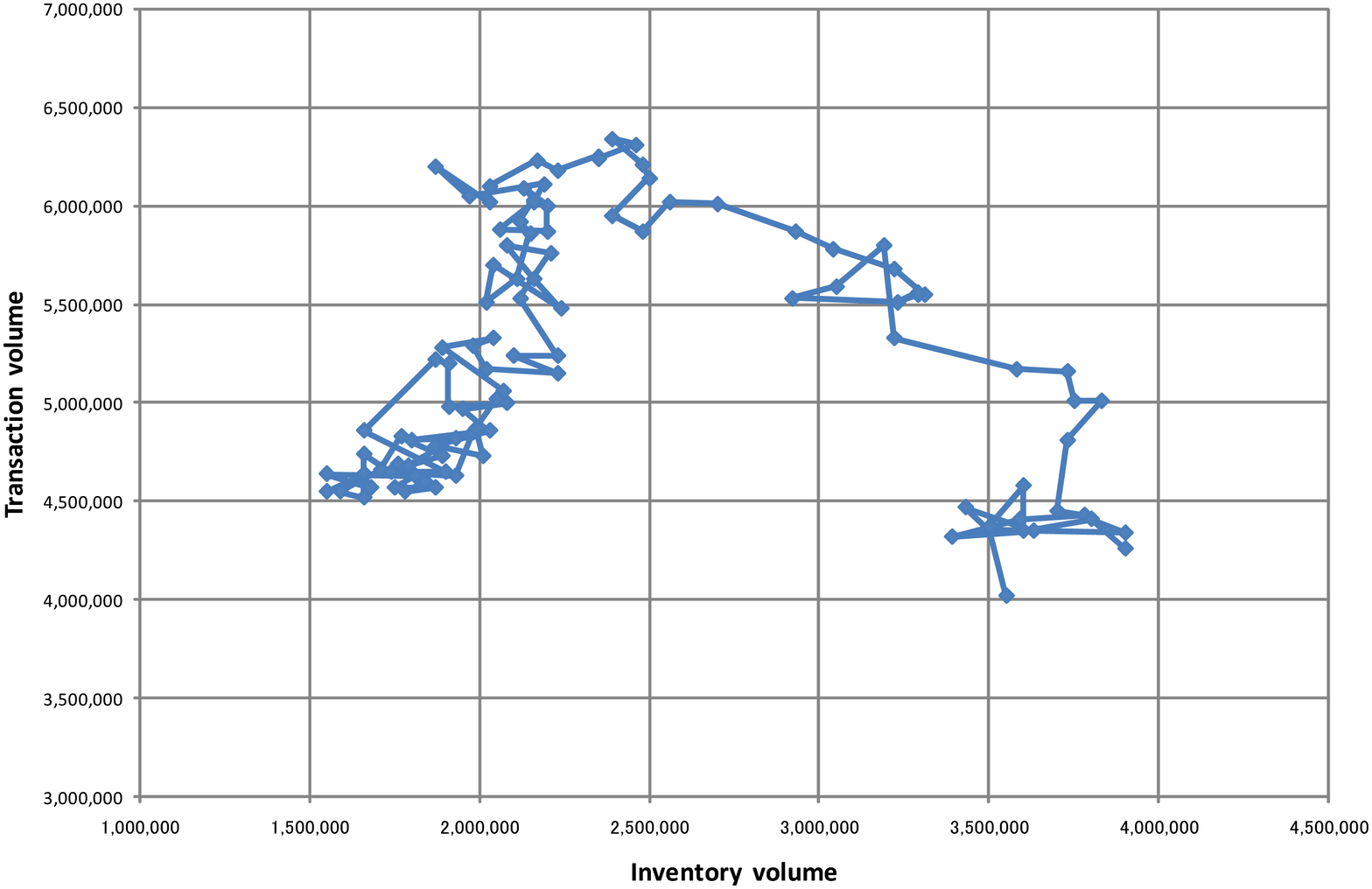}
}
\resizebox{0.49\textwidth}{!}{%
  \includegraphics{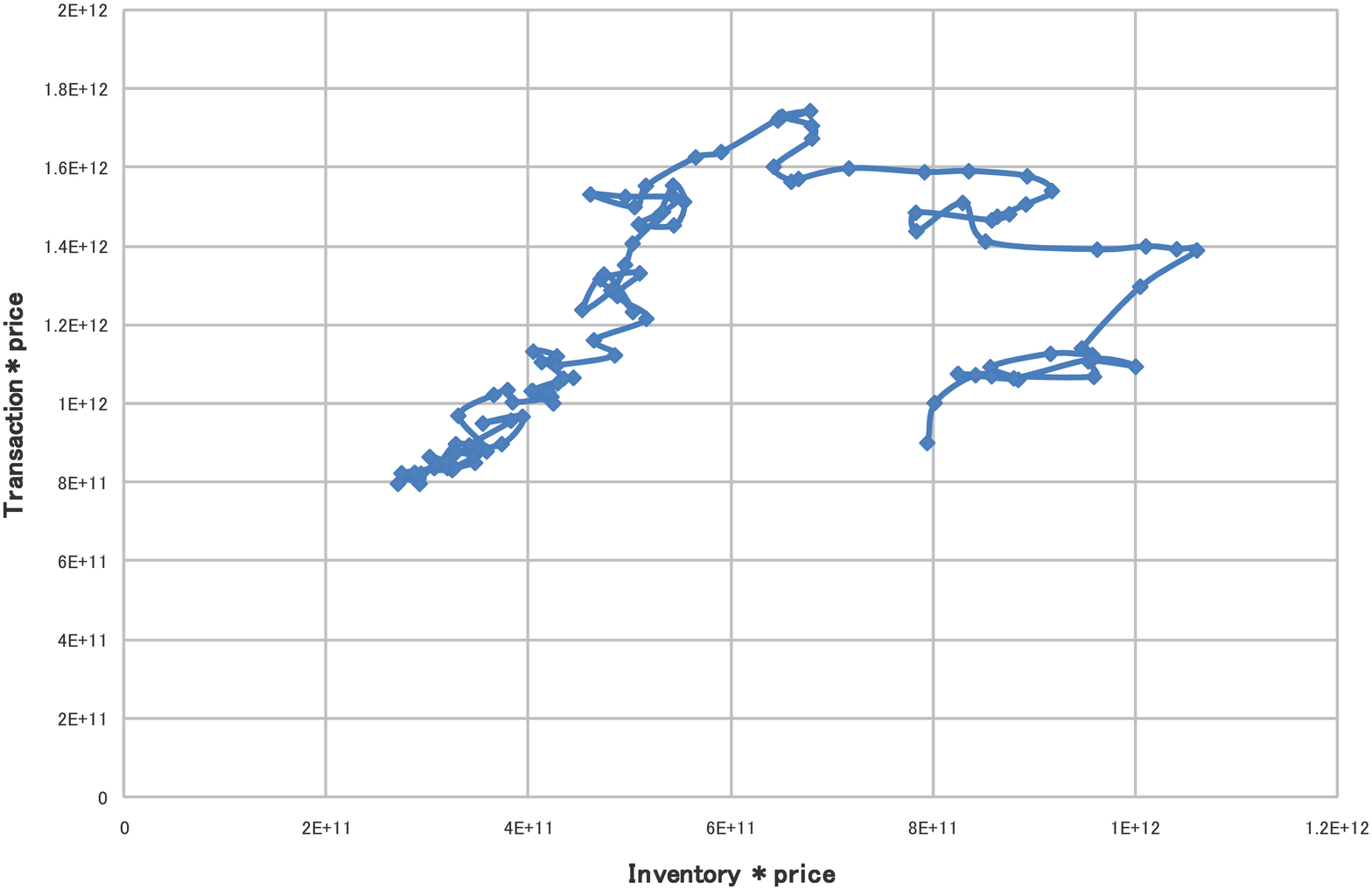}
}
\caption{Fundamental diagram, i.e.,
 inventory versus transaction volume depicted from
 the data shown in Fig.\ref{fig8}. The horizontal and vertical axes represent the number of housing units in the left figure and dollars in the right figure.}
\label{fig9}
  \end{center}
\end{figure}

%%%%%%%%%%%%%%%%%%%%%%%%%%%%%%%%%%%%%%%%%%%%%%%%%%%%%%%%%%%%%%%%%%%%
The variation of transaction volume and inventory is given in Fig.\ref{fig8}.
We see that they both increase first, but at the middle of 2006
transaction begins to fall and inventory increases.
%%%%%%%%%%%%%%%%%%%%%%%%%%%%%%%%%%%%%%%
%%%%%%%%%%%%%%%%%%%%%%%%%%%%%%%%%%%%%%%
Fig.\ref{fig9} is the fundamental diagram of this data, i.e., we depict it
by taking inventory as the horizontal axis and transaction volume as the
vertical axis. Because we cannot observe the total amount of
liquidity supplied to the market, we make two alternative
assumptions in depicting the figures.
The upper one is that the total amount of liquidity changes
in proportion to changes in housing prices.
This represents a situation, for example, that housing prices
continue to rise, and investors have an optimistic expectation
about the future course of prices, and thus willingly provide
additional liquidity to the market.
This assumption is adopted in the upper figure where transaction
and inventory are both measured in the number of housing units.
Alternatively, we may assume that the total liquidity supplied
to the market from the outside investors is
exogenously determined (i.e. it does not depend on the
evolution of prices); more specifically,
we assume that the total amount of liquidity does not change
at all during the entire sample period.
This assumption is adopted in the lower
figure where transaction and inventory are evaluated in dollars.
The two figures should be compared with Fig.\ref{fig7}, showing similar
properties such as increases of flow at early stage
and its collapse due to the decrease of transaction.

%%%%%%%%%%%%%%%%%%%%%%%%%%%%%%%%%%%%%%%
%%%%%%%%%%%%%%%%%%%%%%%%%%%%%%%%%%%%%%%
%%%%%%%%%%%%%%%%%%%%%%%%%%%%%%%%%%%%%%%%%%%%%%%%%%%%%%%%%%%%%%%%%%%%
\section{Concluding discussions}
\label{sec4}
%%%%%%%%%%%%%%%%%%%%%%%%%%%%%%%%%%%%%%%%%%%%%%%%%%%%%%%%%%%%%%%%%%%%
In this paper, we have proposed a bubble burst model
by using a traffic model that represents spontaneous traffic jam.
We find that the phenomenon of bubble burst shares many similar
properties with traffic jam formation.
Especially we would like to stress on the importance of the SlS rule which will
be recast as hesitation in our model.
The simulation results obtained in our model is similar to those of the data
taken from the U.S. housing market.
We focus on transaction in our model instead of the price, and
this has not been considered up to now in studying bubble burst
phenomena. Our result suggests that the transaction could be
a driving force of bursting phenomenon.

In our theoretical model we know the value of the critical density,
then we can predict whether
the market is now in a dangerous metastable state or not by checking
the amount of inventory.
Of course we don't know this critical value in the real market, we hope
that our analysis will help to judge the danger and avoid the burst and
crash of market.
Studying other
bubble phenomena, e.g., oil market and
extensions of our model such as introducing price change and
various dealing network are ongoing and future works.

%%%%%%%%%%%%%%%%%%%%%%%%%%%%%%%%%%%%%%%%%%%%%%%%%%%%%%%%%%%%%%%%%%%%

%%%%%%%%%%%%%%%%%%%%%%%%%%%%%%%%%%%%%%%%%%%%%%%%%%%%%%%%%%%%%%%%%%%%
\end{document}